\documentclass[aps,prl,twocolumn,preprintnumbers,nofootinbib,superscriptaddress]{revtex4-2}

\usepackage{amsmath, mathrsfs, amssymb,amsfonts,amsthm,graphicx, epsf, dcolumn, yfonts, subfigure}
\usepackage[hyperfootnotes=true]{hyperref}
\usepackage{color}
\usepackage{slashed}
\usepackage{setspace}
\usepackage{cancel}
\usepackage{wasysym}
\usepackage{float}
\usepackage{physics}
\usepackage{braket}
\usepackage[normalem]{ulem}
\usepackage[utf8]{inputenc}
\usepackage{leftindex}

\usepackage{color}
\usepackage{hyperref}

\begin{document}


\title{Analytic Tools for Harvesting Magic Resource in Curved Spacetime }

\author{Jiayue Yang}
\email{j43yang@uwaterloo.ca}
\affiliation{Department of Applied Mathematics, University of Waterloo, Waterloo, Ontario N2L 3G1, Canada}
\affiliation{Institute for Quantum Computing, University of Waterloo, Waterloo, Ontario N2L 3G1, Canada}
\affiliation{Perimeter Institute for Theoretical Physics, 31 Caroline St. N., Waterloo, Ontario N2L 2Y5, Canada}
\affiliation{Department of Physics and Astronomy, University of Waterloo, Waterloo, Ontario N2L 3G1, Canada}
\author{Dyuman Bhattacharya}
\email{dyumanb@gmail.com}
\affiliation{Department of Physics and Astronomy, University of Waterloo, Waterloo, Ontario N2L 3G1, Canada}
\author{Ming Zhang}
\email{mingzhang@jxnu.edu.cn }
\affiliation{Department of Physics and Astronomy, University of Waterloo, Waterloo, Ontario N2L 3G1, Canada}
\affiliation{Perimeter Institute for Theoretical Physics, 31 Caroline St. N., Waterloo, Ontario N2L 2Y5, Canada}
\affiliation{Department of Physics, Jiangxi Normal University, Nanchang 330022, China}

\author{Robert B. Mann}
\email{rbmann@uwaterloo.ca}
\affiliation{Department of Applied Mathematics, University of Waterloo, Waterloo, Ontario N2L 3G1, Canada}
\affiliation{Institute for Quantum Computing, University of Waterloo, Waterloo, Ontario N2L 3G1, Canada}
\affiliation{Perimeter Institute for Theoretical Physics, 31 Caroline St. N., Waterloo, Ontario N2L 2Y5, Canada}
\affiliation{Department of Physics and Astronomy, University of Waterloo, Waterloo, Ontario N2L 3G1, Canada}

\begin{abstract}\vspace{-2mm}

The quantum vacuum is not really empty; it is a reservoir of operationally accessible non-classical resources. Understanding how to extract these resources to fuel information processing is a core objective in quantum technologies and lies at the heart of relativistic quantum information (RQI). While earlier studies of quantum resource harvesting protocols relied primarily on numerical methods, we present, for the first time, exact analytic results for the transition probability and coherence of a qutrit Unruh-DeWitt detector interacting with a scalar field in anti-de Sitter  spacetime of arbitrary dimension. Leveraging these results, we analytically investigate  the harvesting of non-stabilizerness and demonstrate that stronger spacetime curvature and higher dimensionality  significantly suppress the amount of extractable  magic resource from the vacuum. Our analytic framework is readily applicable to other scenarios, laying the groundwork for further analytic studies in  RQI.

\end{abstract}

\maketitle

\textit{Introduction---}Understanding the deep connections between quantum physics, information theory, and gravity is one of the most intriguing questions of modern physics,  driving research in areas as
diverse as holography \cite{PhysRevLett.96.181602,VanRaamsdonk:2010pw,Susskind:2014moa,arxiv:1403.5695,arXiv:1406.2678,arXiv:1509.07876,Jefferson:2017sdb,Chapman:2018hou,arXiv:1610.02038,arXiv:2111.02429,Takayanagi:2025ula}, analogue gravity \cite{Unruh:1980cg,Visser:2001fe,barcelo2011analogue,Philbin:2007ji,Wilson_2011,Steinhauer:2015saa,Howl:2016ryt,Jacquet:2020bar} and relativistic quantum information  (RQI) \cite{Peres:2002wx,Fuentes-Schuller:2004iaz,Terno:2005pb,Ball:2005xa,Cliche:2009fma,Ralph:2012mdp,Hu:2012jr,Benincasa:2012rb,Bruschi:2012uf,Martin-Martinez:2012osl,Alsing:2012wf,Ahmadi:2013pfa,Friis:2018nyl}.  The latter has matured into a vibrant cross-disciplinary field   bridging quantum information and relativity, opening new avenues for quantum information processing \cite{Schumacher:1996dy,PhysRevA.57.120,knill2001scheme,monroe2002quantum,Wendin_2017,Flamini_2019,Horodecki:2009zz,nielsen2010quantum,PhysRevA.93.040301,Simidzija:2019zqj,Kasprzak:2024rzj},  quantum teleportation \cite{Alsing:2003es,Alsing:2003cy,Terno:2005pb,Ge:2007ig,Pan:2008yr,Friis:2012cx,Lin:2015ama}, entanglement degradation \cite{Fuentes-Schuller:2004iaz,Alsing:2006cj,Martin-Martinez:2010yva,Martin-Martinez:2009hfq,Martin-Martinez:2010bcj},   entanglement creation \cite{Ball:2005xa,Fuentes:2010dt,Martin-Martinez:2010upl,PhysRevD.85.081701,Friis:2012ki,Brown:2012pw,Dragan:2011hq,Friis:2012nb}, and relativistic implementations of  quantum gates \cite{Bruschi:2012uf,Bruschi:2012pd,Martin-Martinez:2012osl,Martin-Martinez:2014toa,Bruschi:2013tga,Aspling:2024akd,LeMaitre:2024crk}. One key question 
of interest is the extraction of quantum resources from the vacuum of a quantum field \cite{Hotta:2008uk,Pozas-Kerstjens:2015gta,Perche:2022ykt}.  Resource harvesting protocols transfer non-classical resources -- initially hidden and confined within the field -- to localized detectors, where they become operationally accessible. A remarkable example is entanglement harvesting~\cite{Valentini1991,Reznik:2002fz,Reznik:2003mnx,PhysRevA.75.052307,Salton:2014jaa}, in which detectors can acquire entanglement through their interaction with a quantum field, even when they are causally disconnected.

Following these foundational studies, entanglement extraction has been extensively investigated in diverse gravitational settings, such as anti–de Sitter (AdS) spacetime \cite{Henderson:2018lcy,Ng:2018drz}, expanding universes \cite{VerSteeg:2007xs,Nambu:2013rta,Martin-Martinez:2014gra,Kukita:2017etu}, black hole geometries \cite{Henderson:2017yuv,Robbins:2020jca,Gallock-Yoshimura:2021yok,Membrere:2023vao}, and cosmic string spacetime \cite{Ji:2024fcq} (see also \cite{Olson:2010jy,Louko:2006zv,Bruschi:2010mc,Hodgkinson:2011pc,PhysRevD.90.064003,Sabin:2012pj,Martin-Martinez:2015qwa,Beny:2017wwe,Henderson:2020zax,Liu:2020jaj,Zhang:2020xvo,Liu:2021dnl,Patterson:2022ewy,Suryaatmadja:2022quq,Preciado-Rivas:2024gzm,Wang:2025lga} for other  circumstances). However, in curved spacetime all   existing studies rely on numerical integration (or multiple sums over field modes), with analytic results for the detector’s density matrix elements lacking. In this Letter, we break new ground by presenting analytic derivations and results of key quantities 
such as transition probability and coherence.  We demonstrate our analytic methods by computing the extraction
of non-stabilizerness (or magic resource) \cite{Gottesman:1997qd,Gottesman:1998hu,Gottesman:1999tea,Aaronson:2004xuh,BravyiK05} for a qutrit detector~\cite{LimaEtAl23} in $d+1$-dimensional AdS spacetime, a maximally symmetric manifold with constant negative curvature that serves as a cornerstone in quantum gravity and holography~\cite{tHooft:1993dmi, Susskind:1994vu, Maldacena:1997re, Gubser:1998bc, Witten:1998qj}. Note that our analytic tools also apply to entanglement and magic harvesting in various scenarios. To the best of our knowledge,  this is the first instance in which exact analytic results of this kind 
  have been obtained. An advantage of our analytic approach is the straightforward renormalization of divergences that would otherwise hinder reliable numerical evaluation.  Furthermore, as the dimension gets larger ($d>4$) numerical methods \cite{Henderson:2018lcy,Ng:2018drz} become less precise and eventually break down, whereas  our method is applicable 
  for arbitrary $d$, yielding   deeper insight into the resource harvesting protocol and opening up a new set of analytic tools for RQI.

Non-stabilizerness, commonly referred to as magic resource, quantifies the non-Clifford component required to achieve  universal fault-tolerant quantum computation and  encapsulates the capability of a quantum system to  perform tasks that are classically intractable~\cite{Gottesman:1997qd,Gottesman:1998hu,Gottesman:1999tea,Aaronson:2004xuh,BravyiK05}.
Although entanglement characterizes non-classical correlations, entangled states can still be efficiently
 simulated classically, indicating that entanglement alone does not necessarily signify the quantum advantage~\cite{Gottesman:1997qd,Gottesman:1998hu,Gottesman:1999tea,Aaronson:2004xuh,BravyiK05}.  Unlike entanglement, magic resource acts as a more fundamental indicator of quantum advantage and  plays a central role in quantum computation.  Recent advances have revealed that the magic resource is more than just computational power. It has been employed to diagnose criticality in many-body physics \cite{Sarkar_2020,White:2020zoz,PRXQuantum.3.020333,Catalano:2024bdh,Hoshino:2025jko}, quantify complexity in nuclear physics and particle physics \cite{PhysRevResearch.7.023228,Brokemeier:2024lhq,Robin:2024bdz}, describe  information scrambling and  retrieval in black holes \cite{Hayden:2007cs,PhysRevLett.105.203901,Yoshida:2017non,PhysRevA.106.062434,Garcia:2022abt,Liu:2025qfl},   and shed light on  gravitational back-reaction \cite{Cao:2024nrx} as well as the emergence of spacetime \cite{Goto:2021anl}. From the RQI perspective, a central issue is whether the vacuum can supply such quantum computational power. While entanglement harvesting probes the vacuum’s entanglement structure, magic harvesting provides a complementary probe of the vacuum’s intrinsic computational resource. Operationally, the magic harvesting protocol offers a vision for future quantum computers: Clifford operations are
performed natively, while ancillary qudits coupled to a quantum field harvest magic resource from the vacuum on
demand, supplying the fuel needed to power universal quantum computation \cite{Nystrom:2024oeq}.  While recent studies have extracted magic resource~\cite{Nystrom:2024oeq,Cepollaro:2024sod} (and contextuality~\cite{LeMaitre:2025fnt,Lima:2025ton}) in flat Minkowski space~\cite{Nystrom:2024oeq,Cepollaro:2024sod,LeMaitre:2025fnt,Lima:2025ton}, no studies have considered magic resource harvesting in curved spacetimes -- despite gravitational effects and lab-induced effective curvature being inevitable in practice \cite{Szpak:2014jua,Tononi_2024,Huhtala:2001cj,Stegmann:2015mjp,Yang:2024fql}.  Intriguing questions naturally follow.  Does curvature enhance or suppress the availability of non-stabilizerness?  How does the dimensionality of spacetime influence the extraction of magic resource?  In this Letter, we shall employ our exact analytic results to address these fundamental questions, paving the way for a deeper understanding of how quantum resource harvesting manifests in curved spacetime.

\textit{Qutrit Unruh-DeWitt detector in AdS spacetime---}In quantum field theory in curved spacetimes, there is no well-defined concept of a particle or unique vacuum state. As a result, particles are defined operationally—as entities detected by particle detectors, most commonly
  the Unruh-DeWitt (UDW) detector~\cite{Unruh76, DeWitt80},  a two-level quantum mechanical system (i.e., a qubit)  that locally interacts with a quantum field. We shall consider a qutrit UDW detector (rather than a qubit),  a three-level quantum system with quantum states (or computational basis) $\{\ket{0}_{D},\ket{1}_{D}, \ket{2}_{D}\}$ and respective energy
eigenvalues $\{0,\Omega_1 , \Omega_1 + \Omega_2\}$, which  is described by a free Hamiltonian $\hat{H}_{\text{free}}=\text{diag}(0,\Omega_1 , \Omega_1 + \Omega_2)$
\cite{LimaEtAl23}.  In keeping with standard terminology, we still refer to it as a UDW detector. 
The detector moves along a trajectory $\rm{x}(\tau)$ parameterized by its proper time $\tau$   and locally interacts with a real massless quantum scalar field $\hat{\phi}$. The interaction picture Hamiltonian is then given by $ \hat{H}_{\text{int}}(\tau) =\lambda \chi(\tau)\hat{\mu}(\tau)\otimes\hat{\phi}(\rm{x}(\tau))$, where $\lambda$ is the detector--field coupling  strength parameter, $\chi(\tau)$ is a switching function that characterizes the time interval during which the detector-field interaction occurs,  $\hat{\phi}(\rm{x}(\tau))$ is the Hermitian quantum field operator, evaluated along the detector's trajectory, and  $  \hat{\mu}(\tau) \equiv2^{-1/2}\left(\right. \ket{1}_D \leftindex_D{\bra{0}} e^{i\Omega_1 \tau} + \ket{2}_{D}\leftindex_D{\bra{1}} e^{i \Omega_2 \tau})+\text{H.c.}$ is the monopole moment operator that characterizes the state transitions of the qutrit detector~\cite{LimaEtAl23}. 
 
The time evolution of the system during the detector-field interaction is governed by the unitary operator  $\hat{U} := \mathcal{T} \exp\left[ -i \int \mathrm{d}\tau \hat{H}_{\text{int}}(\tau) \right]$ generated by the interaction Hamiltonian, where the time-ordering operator \(\mathcal{T} \) is defined as $    \mathcal{T}  A(t) B(t') := \Theta(t - t') A(t) B(t') + \Theta(t' - t) B(t') A(t)$, 
with $\Theta$ being the Heaviside step function. We consider the initial state of the system to be $ \ket{\psi}=\ket{0}_{D}\otimes\ket{0}$, with corresponding density operator $ \hat{\rho}_0 = \ket{\psi}\bra{\psi}=\ket{0}_{D}\leftindex_D{\bra{0}} \otimes \ket{0}\bra{0}.$
Here $\ket{0}$ denotes the vacuum state of the quantum field, which satisfies $\bra{0}\hat{\phi}\ket{0}=0$. 
The reduced density operator for the final state of the detector, obtained by tracing over the field degrees of freedom, is $\hat{\rho}_\mathcal{D}={\rm Tr}_\phi[\hat{U} \   \hat{\rho}_0\ \hat{U}^\dagger]=\sum_{i,j}\text{Tr}_\phi [\hat{U}^{(i)} \hat{\rho}_0 \hat{U}^{\dagger (j)}]$, where we denote \( \hat{U}^{(i)} \) as the \( i \)-th term in the Dyson perturbative expansion of \( \hat{U} \), which is of order \( \lambda^i \). The reduced density operator, expressed in matrix form with respect to the qutrit's computational basis  $\{\ket{0}_{D},\ket{1}_{D},\ket{2}_{D}\}$, is given by \cite{LimaEtAl23,Nystrom:2024oeq}
\begin{equation}
    \hat{\rho}_\mathcal{D}
    =\begin{bmatrix}
       \ p & \ \ \ 0 & \beta^* \\ \ \  0 \ \ &\ \ q & 0 \\\  \beta & \ \ 0 & 1-p-q
    \end{bmatrix}\,,
    \label{rho2}
\end{equation}
where $p=\leftindex_D{\bra{0}} \hat{\rho}_\mathcal{D} \ket{0}_{D}$ quantifies the probability that the detector is found in its ground state, $q=\leftindex_D{\bra{1}} \hat{\rho}_\mathcal{D} \ket{1}_{D}$ represents the excitation probability to the first excited state $\ket{1}_D$, while the off-diagonal term $\beta=\leftindex_D{\bra{2}}\hat{\rho}_\mathcal{D} \ket{0}_{D} $ encodes the coherence between the ground state $\ket{0}_D$ and the second excited state $\ket{2}_D$. Assuming the detector–field coupling strength parameter \( \lambda \) is sufficiently small for perturbation theory to be valid, we can express \( q \) (to second order) as
\begin{align}
     q 
     &=\frac{ \lambda^2 }{2}\int \mathrm{d}\tau \mathrm{d}\tau' \chi(\tau)\chi(\tau')e^{-i\Omega_1(\tau-\tau')}{\mathcal{W}}(\tau,\tau')\,,\label{qq}
\end{align}
with  $  {\mathcal{W}}(\tau,\tau')\equiv \bra{0}\hat{\phi}(\rm{x}(\tau))\hat{\phi}(\rm{x}(\tau'))\ket{0}$ being the vacuum Wightman function. The off-diagonal element is 
\begin{align}
    \beta =&  -\frac{\lambda^2}{4} \int \mathrm{d}\tau \mathrm{d}\tau' \chi(\tau)\chi(\tau') \nonumber \\ &\times \bigg[ \Theta(\tau-\tau') e^{i(\Omega_1 \tau' + \Omega_2 \tau)} {\mathcal{W}}(\tau, \tau') \nonumber\\
    &\quad+\Theta(\tau'-\tau) e^{i(\Omega_1 \tau + \Omega_2 \tau')}{\mathcal{W}}(\tau', \tau) \bigg]  + \mathcal{O}(\lambda^4)\,.\label{beta}
\end{align}
Consider a static detector in $D=d+1$-dimensional AdS spacetime with coordinates $\left(t, r, \theta_1, \theta_2, \ldots, \theta_{d-1}\right)$, the corresponding line element takes the form
\begin{equation}\label{metric}
d s^2=-\left(1+r^2 / \ell^2\right) \mathrm{d} t^2+\frac{\mathrm{d} r^2}{1+r^2 / \ell^2}+r^2 \mathrm{d} \Omega_{d-1}^2\ , 
\end{equation}
where $\ell$ is the AdS radius, $\mathrm{d} \Omega_{d-1}^2$ is the line element for unit $(d-1)$-dimensional sphere, and the scalar curvature of AdS${}_{d+1}$ is given by $\mathcal{R}=-d(d+1)/\ell^2.$ For a static detector, the trajectory parameterized by the proper time \(\tau\) is $\mathrm{x}(\tau) = \{ t = \frac{\tau \ell}{\gamma}, \; r = R, \; \theta_i = \Theta_i\}$,
where $\gamma = \sqrt{\ell^2+R^2}$, $R$ and $\Theta_i$  are constant parameters that define the detector’s location. The regularized Wightman function   is given by
(see Supplemental Material for derivations) \cite{Decanini:2005gt,Decanini:2005eg,Jennings:2010vk,saharian2020quantum,Smith:2017vle,Pitelli:2021oil,Louko:2014aba}  
\begin{equation}\label{wight}
{\mathcal{W}}(\tau, \tau') = \frac{\Gamma(d-1)}{(4\pi)^{d/2} \Gamma(\frac{d}{2})} \left[ 2 i\gamma \sin \left(\frac{\tau -\tau'-i\epsilon}{2 \gamma}\right)\right]^{1-d}\,,
\end{equation}
where $d$ ($d\geq2$) is the spatial dimension, $\Gamma$ represents the Gamma function, $\epsilon$ is  the regulator responsible for the  $i\epsilon$-regularization in the Wightman function. For the switching function, we choose $\chi(\tau)$ to be a smooth Gaussian function with variance $\sigma^2/2$ given by $ \chi(\tau) = e^{-\tau^2 / \sigma^2}$.

 \textit{Analytic derivations for transition and coherence---}As shown in \eqref{wight}, the Wightman function along the worldline of a static detector depends only on the proper time difference. Thus starting from \eqref{qq}, introducing the variables $u = \tau + \tau'$ and $s = \tau - \tau'$, and  letting $\Omega_1=\Omega_2=\Omega$ for simplicity, we arrive at 
\begin{align}
    q =& \frac{\lambda^2}{4}\int_{-\infty}^{\infty}  \mathrm{d}u\; \int_{-\infty}^{\infty} \mathrm{d}s\; e^{-(u^2+s^2)/2\sigma^2}e^{-i\Omega s}{\mathcal{W}}(s)\nonumber\\
    =& \frac{\lambda^2}{2}\sigma\sqrt{\frac{\pi}{2}} \int_{-\infty}^{\infty}\mathrm{d}s\; e^{-s^2/2\sigma^2}e^{-i\Omega s}{\mathcal{W}}(s)\,.
\end{align}
Substituting the regularized Wightman function \eqref{wight}, with $1/2\gamma$ absorbed into $\epsilon$, yields 
\begin{align}
    q 
    =& \frac{\lambda^2\sigma\sqrt{\pi/2} \gamma^{1-d}\Gamma(d-1)}{2(4\pi)^{d/2} \Gamma(\frac{d}{2})}\int_{-\infty}^{\infty} \mathrm{d}s\;e^{-s^2/2\sigma^2} \nonumber\\
    & \times e^{-i\Omega s}\left[2i\sin \left(\frac{s}{2\gamma}-i\epsilon\right)\right]^{1-d}\,.
\end{align}
Applying 
the generalized binomial theorem, we have \begin{align}\label{jfweiio}
&\left[2i\sin\left(\frac{s}{2\gamma} -i\epsilon \right)\right]^{1-d}=\left[e^{i(\frac{s}{2\gamma}-i\epsilon)}-e^{-i(\frac{s}{2\gamma}-i\epsilon)}\right]^{1-d}\nonumber\\
=&e^{(1-d)i(\frac{s}{2\gamma}-i\epsilon)}\left[1-e^{-2i(\frac{s}{2\gamma}-i\epsilon)}\right]^{1-d}\nonumber\\
=&\sum_{n=0}^{\infty}\binom{d+n-2}{n}e^{-i(2n+d-1)(\frac{s}{2\gamma}-i\epsilon)}\,.
\end{align}
Crucially, the choice of which exponential to factor out is nontrivial, since $(1-x)^{-(d-1)}$ only converges for $|x|<1$, here 
 ensured since $\epsilon > 0$. Thus
 (see Supplemental Material for details)
\begin{align}
    q=&\frac{\lambda^2\sigma\sqrt{\pi/2}\gamma^{1-d}\Gamma(d-1)}{2(4\pi)^{d/2} \Gamma(\frac{d}{2})}\sum_{n=0}^{\infty}\binom{d+n-2}{n}e^{-(2n+d-1)\epsilon}\nonumber\\
    &\int_{-\infty}^{\infty} \mathrm{d}s\; e^{-s^2/2\sigma^2}e^{-i\Omega s}e^{-i(2n+d-1)s/2\gamma}  \nonumber\\
    =& 2\alpha\sum_{n=0}^{\infty}\frac{\Gamma(d+n-1)}{\Gamma(n+1)}e^{-\frac{\sigma^{2}}{2}\left(\Omega+\Omega_n\right)^{2}}\,,
    \label{qexact}
\end{align}
where $\alpha\equiv\frac{\lambda^2\sigma^{2}\pi\gamma^{1-d}}{4(4\pi)^{d/2} \Gamma(\frac{d}{2})}$ and $\Omega_n\equiv\frac{2n+d-1}{2\gamma}$.
\eqref{qexact} provides an exact analytic expression for the transition probability $q$, obtained by
integrating over $s$, taking the limit $\epsilon\rightarrow0^{+}$,  and rewriting the binomial coefficient in terms of the Gamma function $\Gamma$.   

To compute  $\beta$ from \eqref{beta}, we again use the $(u,s)$
variables, obtaining
\begin{align}
    \beta=&-\frac{\lambda^2}{8} \int_{-\infty}^{\infty} \mathrm{d}u\; \int_{-\infty}^{\infty}\mathrm{d}s\; \bigg[ e^{-(u^2 + s^2)/2\sigma^2} e^{i\Omega u} \nonumber\\
    &\times \left(\Theta(s){\mathcal{W}}(s) + \Theta(-s){\mathcal{W}}(-s)\right) \bigg]\nonumber\\
     =&-\frac{\lambda^2\sqrt{2\pi}\sigma}{4}e^{-\sigma ^2 \Omega ^2/2} \int_{0}^{\infty} \mathrm{d}s\; e^{-s^2/2\sigma^2}{\mathcal{W}}(s)\,,
\end{align}
yielding in turn (see Supplemental Material for  details)
\begin{align}\label{beta51main}
\beta =&-\alpha \sum_{n=0}^{\infty}\frac{\Gamma(d+n-1)}{\Gamma(n+1)}e^{-\frac{\sigma^{2}}{2}\left(\Omega^2+\Omega_n^2\right)}\,,
\end{align}
which is 
the explicit analytic expression for the renormalized coherence term $\beta$. 

 The exact series expressions for the transition probability $q$ in \eqref{qexact}  and coherence $\beta$ in \eqref{beta51main} reveal a rich underlying structure that offers deep physical insight. These results can be understood as a superposition of contributions from individual modes, modulated by spacetime dimension and curvature.  Each term in the summation corresponds to a discrete excitation mode labeled by \( n \), with its contribution  weighted by  a Gamma-function prefactor and exponentially damped by a Gaussian function.  While the expression takes the form of an infinite sum, the exponential term imposes a fast convergence and  natural spectral cutoff, sharply suppressing high-$n$ contributions and indicating that the series is dominated by low-lying modes. Notably, 
 the analytic expressions for $q$  and $\beta$ have a similar  structure. This becomes evident once we introduce an effective energy level $\tilde{\Omega}_n$ that is discrete and mode-dependent.  For $q$, we define $\tilde{\Omega}_n^{(q)} = \Omega + \Omega_n$, while for $\beta$, it takes the form $\tilde{\Omega}_n^{(\beta)} = \sqrt{\Omega^2 + \Omega_n^2}$. Note that $\Omega_n=\frac{2n+d-1}{2\gamma}$, so the effective energy level $\tilde{\Omega}_n$  inherently encodes the effects of curvature and dimensionality of the spacetime through its dependence on $\gamma$ and $d$. Its discreteness, moreover, may hint at underlying quantum effects.  In the Minkowski limit $\ell\to \infty$, $\tilde{\Omega}_n$ naturally reduces to the detector’s intrinsic energy gap $\Omega$. 
 
 Remarkably, we find the series $q$  and $\beta$  admit closed-form expressions $q_c$ and $\beta_c$ in the large $\gamma$ regime as (see Supplemental Material)
\begin{align}\label{closeqmain}
   q_c=& 2\gamma \alpha\sum_{n=0}^{d-2} 
\zeta_n^{(q)}\left[\Gamma \left(\frac{n+1}{2}\right)  F\left(-\frac{n}{2};\frac{1}{2};\frac{- \sigma ^2 \Omega'^2}{2}\right)\right.\nonumber\\
&\left.-\sqrt{2} \sigma\Omega' \Gamma \left(\frac{n+2}{2}\right)  F\left(\frac{1-n}{2};\frac{3}{2};\frac{-\sigma ^2 \Omega'^2}{2}\right)\right]\,,\nonumber\\
   \beta_c=& -\gamma \alpha\sum_{n=0}^{d-2} 
\zeta_n^{(\beta)}\left[\Gamma \left(\frac{n+1}{2}\right)  F\left(\frac{n+1}{2};\frac{1}{2};\frac{\Omega_d^2 \sigma ^2}{2}\right)\right.\nonumber\\
&\left.-\sqrt{2} \sigma  \Omega_d \Gamma \left(\frac{n+2}{2}\right)F\left(\frac{n}{2}+1;\frac{3}{2};\frac{\Omega_d^2 \sigma ^2}{2}\right)\right]
\end{align}
where $\Omega'=\Omega+\Omega_d$,  $\Omega_d= \frac{d-1}{2\gamma}$, 
$F(a;b;c)$ represents the Kummer confluent hypergeometric function, $\zeta_n^{(q)}= 2^{\frac{n-1}{2}}\sigma ^{-n-1}\mathcal{C}_n^{(d)} $, $\zeta_n^{(\beta)} =\zeta_n^{(q)} e^{-\frac{1}{2} \sigma ^2 \left(\Omega^2+\Omega_d ^2\right)}$, and  $ \mathcal{C}_n^{(d)}$
is the coefficient of  $x^n$ in $ \prod_{k=1}^{d-2} (\gamma x + k)
$. Furthermore, one can analytically verify that their forms in AdS${}_{d+1}$ spacetime reduce to those in Minkowski spacetime as $\gamma\to \infty$.

\textit{Exact results for mana and the mana discriminant---}We now apply our exact analytic results to the non-stabilizerness harvesting protocol. Several measures have been developed to quantify non-stabilizerness, or magic resource, including Wigner negativity \cite{Emerson:2013zse}, stabilizer rank \cite{Bravyi:2018ugg}, stabilizer Rényi entropy  \cite{Leone:2021rzd}, and mana~\cite{VeitchETAl14}. In this Letter, we adopt mana—a well-established and operationally significant measure—to quantify magic resource. In the interaction picture of the energy eigenbasis, mana $  M(\hat{\rho}_\mathcal{D})$ is determined by the elements $q$ and $\beta$ of the detector's final-state density matrix \eqref{rho2} \cite{Nystrom:2024oeq},
\begin{align}
    M(\hat{\rho}_\mathcal{D}) =& \ln\left[ 1-q + \frac{1}{3} \left( \abs{q-\textrm{Re}(\beta)-\sqrt{3}~\textrm{Im}(\beta)}\right.\right.\nonumber\\
    & \left.\left. + \abs{q+2~\textrm{Re}(\beta)} + \abs{q-\textrm{Re}(\beta)+\sqrt{3}~\textrm{Im}(\beta)} \right)\right]~.
\end{align}

Interestingly, we can identify a simple but powerful diagnostic $\Delta:=-(q+2\beta)$, analogous to the discriminant in a quadratic equation, that determines whether magic resource can be harvested.
When the diagnostic $\Delta\leq0$, magic resource extraction is forbidden; otherwise,  magic resource  harvesting  becomes feasible and the amount of mana that can be harvested  becomes $ M_{\Delta}=  \ln \left(1+2\Delta/3\right)$, which is determined solely by $\Delta$. This ``mana discriminant'' offers a clear and intuitive criterion for diagnosing quantum resource accessibility in RQI.   Note that the proposed diagnostic applies in the regime where $q$ is real and positive (as is automatically satisfied by any physical excitation probability), and $\beta$ is real and negative. Extending this criterion to the more general case with complex $\beta$ warrants further investigation. Using \eqref{qexact} and
\eqref{beta51main} we obtain
\begin{align}
\label{Deltexp}
    \Delta
=&2\alpha\sum_{n=0}^{\infty}\frac{\Gamma(d+n-1)}{\Gamma(n+1)}e^{-\frac{\sigma^{2}}{2}\left(\Omega^2+\Omega_n^2\right)}\nonumber\\
&\times \left(1-e^{-\sigma^{2}\Omega\Omega_n}\right) >0
\end{align}
since
 $1-e^{-\sigma^{2}\Omega\Omega_n}>0$ for a positive energy gap and all  other factors are positive.  Hence we obtain the general result that the magic resource can be harvested in AdS spacetime. 

\begin{figure}
\includegraphics[width=8.5cm]{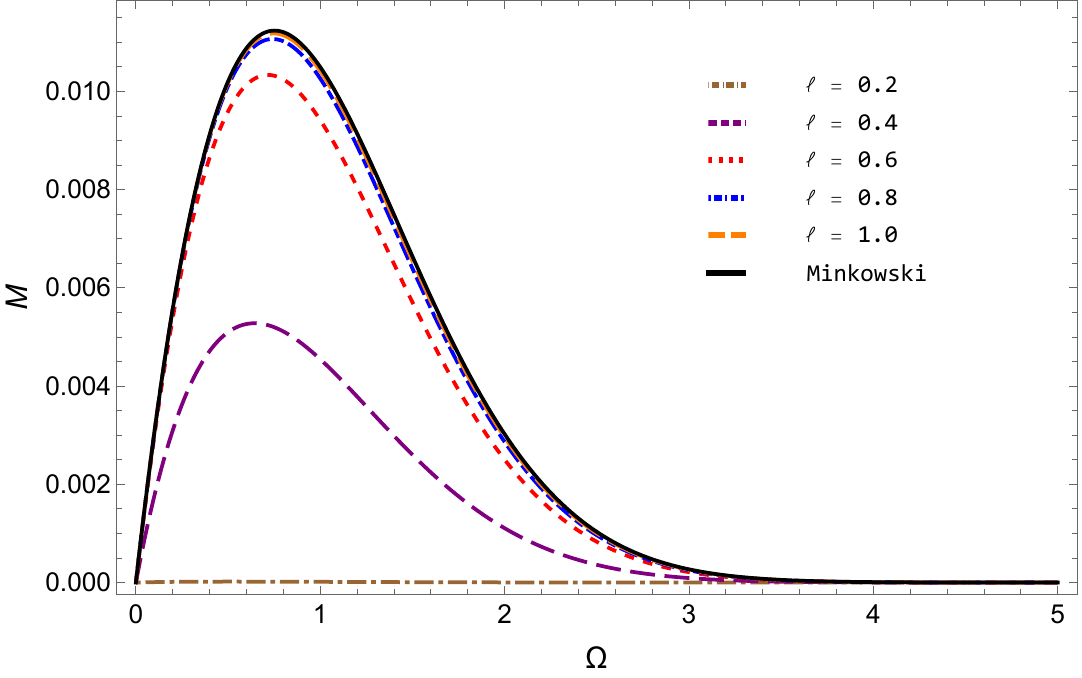}
\caption{Diagram of  mana  \( M \) versus energy gap \( \Omega \) in  AdS${}_{4}$   with different AdS length scales $\ell$. The detector is set to be static at $R=0.1$,  and we also set $\sigma=1$  and $\lambda=1$. \label{AdS4}}
\end{figure}

\begin{figure}
\includegraphics[width=8.5cm]{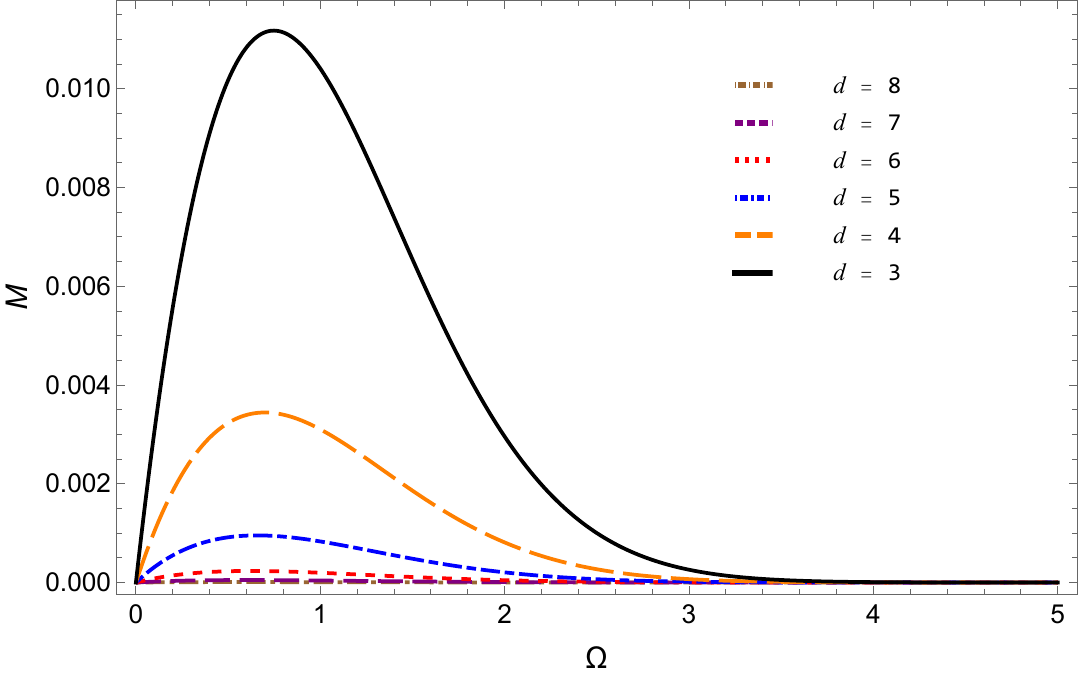}
\caption{Diagram of  mana  \( M \) versus energy gap \( \Omega \) in AdS${}_{d+1}$ with different dimensions. The detector is set to be static at $R=0.1$, and we also set  $\sigma=1$, $\lambda=1$, and  $\ell=1$. \label{AdSn}}
\end{figure}

\textit{Effects of curvature and dimensionality on magic resource harvesting---}In Figs. \ref{AdS4} and~\ref{AdSn}, we present the mana \( M \) harvested by a static detector at \( R = 0.1 \) in AdS spacetime, as a function of the energy gap $\Omega$, for various AdS radii  and spacetime dimensions. The behavior of mana in various geometries reveals a striking feature. It exhibits a non-monotonic dependence on the energy gap $\Omega$ with a distinct peak across all backgrounds. This  behavior  
can be directly understood from our analytic expression
\eqref{Deltexp}.  For small $\Omega$, the analytic structure of $\Delta$ reveals a linear dependence on $\Omega$, leading to $M_\Delta \sim \Omega$. This reflects the initial rise in the  extractable mana. At large $\Omega$, the explicit form yields $M_\Delta \sim e^{-\frac{\sigma^2}{2} \Omega^2}$, signaling the exponential decay of the mana as 
the energy gap increases.  The initial linear growth and eventual exponential damping lead to a pronounced peak in $M_\Delta(\Omega)$, indicating an energy gap at which the detector is optimally tuned to harvest magic resource. This may suggest a resonance-like phenomenon where the detector’s energy gap matches the field modes, leading to maximal extraction of nonclassicality.

Notably, in Fig. \ref{AdS4}, as the AdS curvature becomes stronger (i.e., as $\ell $  decreases), the mana peak shifts toward lower values of the energy gap $\Omega$. This reflects a curvature-induced shift in the field–detector resonance structure, or a potential manifestation of gravitationally induced ``redshift'' in the context of RQI. Furthermore, the AdS curvature plays a crucial role in determining the amount of magic resource that can be harvested. As $\ell$ decreases, AdS spacetime becomes more strongly curved, and the mana is significantly suppressed. In the ultra-strong curvature regime (i.e., $ \ell \le 0.2 $), the suppression becomes so severe that nearly no mana can be extracted, potentially signaling an effective decoupling of the detector and the field, and a curvature-induced “freezing” of magic resource harvesting. At large $\ell$, the mana harvested asymptotically approaches the flat-spacetime (Minkowski) result, confirming the validity of our formalism under varying spacetime curvature. The effect of AdS curvature on magic resource harvesting can be traced to the analytic structure of $\Delta$ \eqref{Deltexp}. As the AdS radius $\ell$ decreases, $\gamma = \sqrt{\ell^2 + R^2}$ becomes smaller, resulting in larger $\alpha = \frac{\lambda^2 \sigma^2 \pi \gamma^{1-d}}{4(4\pi)^{d/2} \Gamma\left(\frac{d}{2}\right)}$ and larger $\Omega_n = (2n + d - 1)/(2\gamma)$. Although larger $\alpha$ tends to amplify $\Delta$, this effect is subdominant compared to the exponential suppression from $e^{-\frac{\sigma^2}{2}(\Omega^2 + \Omega_n^2)}$. As a result, the peak of $M_\Delta$ becomes lower and shifts leftward.  In the strong curvature limit $\ell \to 0$, the suppression becomes overwhelming and the mana vanishes.

We now investigate the impact of spacetime dimensionality on magic resource harvesting.  As shown in Fig. \ref{AdSn}, the mana displays a similar pattern across different dimensions in AdS$_{d+1}$: a rise to a peak followed by a drop, indicating that the resonance-like behavior is a generic feature of our magic resource harvesting protocols. However, the magnitude of mana that can be harvested exhibits a pronounced dependence on the dimensionality of AdS${}_{d+1}$  spacetime.  As the dimension increases, the overall harvested mana is significantly suppressed, and the peak shifts to smaller energy gap values, indicating that the optimal $\Omega$ value occurs at lower frequencies in higher dimensions. The analytic structure of \eqref{Deltexp} also offers insight into the dimension-dependent behavior of extractable magic resource. Since $\Omega_n$ increases with $d$, the exponential suppression $e^{-\frac{\sigma^2}{2}(\Omega^2 + \Omega_n^2)}$ becomes stronger at higher dimensions, resulting in a lower and left-shifted peak in $M_\Delta$. This suggests that lower-dimensional AdS spacetimes are  more favorable for extracting quantum resources from the vacuum. 

\textit{Discussion---}As the first study to derive exact series expressions for the excitation probability and off-diagonal coherence of a UDW detector in RQI, this Letter unveils the underlying analytic structure of these quantities, marking an essential step toward a more complete analytic understanding of RQI. Our results shift the focus from predominantly numerical approaches to exact analytic series solutions, establishing a new paradigm for studies in RQI. Although we use magic resource harvesting in AdS${}_{d+1}$ spacetime as an example to illustrate our analytic methods, the applicability of our methods extends far beyond the present context; the analytic framework can be efficiently applied to other scenarios, such as the entanglement harvesting protocol,  where the transition probability and coherence in the density matrix of two-qubit detectors exhibit structural similarities to those of the single-qutrit case analyzed here.

As the first exploration of magic resource harvesting in curved spacetimes, our work offers a novel perspective on how quantum resource harvesting is shaped by gravitational effects. We demonstrate that both spacetime curvature and dimensionality act as geometric modulators of extractable quantum resources: as curvature increases and dimensionality rises, the amount of magic resource that can be harvested from the gravitational vacuum diminishes, and the optimal energy gap shifts downward. While our analysis focuses on AdS spacetime, the analytic framework we develop can be readily extended to other gravitational backgrounds. Our findings open new avenues for analytically probing the intricate relationship between spacetime geometry and quantum resource harvesting and are expected to inspire further investigations into the deep interplay between gravitational theory and quantum information theory.

 Overall, the analytic framework developed here offers a versatile toolkit for RQI, with promising applications to quantum resource harvesting in a wide range of scenarios. In light of the central role of AdS spacetime in holography, our results may also shed light on the quantum nature of spacetime, contributing to foundational questions at the intersection of quantum information and quantum gravity.

\textit{Acknowledgements---}We thank Sanchit Srivastava for inspiring talks and discussions,  María Rosa Preciado-Rivas, Gong Cheng and Shiyue Ren for helpful discussions, and the  anonymous referees for their insightful comments. This work was supported in part by the Natural Sciences and Engineering Research Council of Canada. The work of MZ received support from National Natural Science Foundation of China (Grant No. 12365010) and Chinese Scholarship Council Scholarship. Research at Perimeter Institute is supported in part by the Government of Canada through the Department of Innovation, Science, and Economic Development and by the Province of Ontario through the Ministry of Colleges and Universities.

\bibliographystyle{apsrev4-2} 
\bibliography{refs}

\newpage

\onecolumngrid

\section{Supplemental material}

\subsection{Wightman Function in $\text{AdS}_{d+1}$ Spacetime}
As a solution to Einstein’s equations with a negative cosmological constant $\Lambda = -\frac{d(d - 1)}{2 \ell^2}$, anti-de Sitter (AdS) spacetime is a maximally symmetric spacetime characterized by constant negative curvature. In particular, the $d+1$-dimensional AdS spacetime, can be represented as a  hyperboloid
\begin{equation}
\left(z^0\right)^2-\left(z^1\right)^2-\cdots-\left(z^d\right)^2+\left(z^{d+1}\right)^2=\ell^2\,,
\end{equation}
embedded in a flat $(d+2)$-dimensional background
\begin{equation}
\mathrm{d} s^2=-\left(\mathrm{d} z^0\right)^2+\left(\mathrm{d} z^1\right)^2+\cdots+\left(\mathrm{d} z^d\right)^2-\left(\mathrm{d} z^{d+1}\right)^2\,,
\end{equation}
with $z^0$ and $z^{d+1}$ being time-like coordinates , and $z^i (i=1,\cdots,d)$ being space-like coordinates, the parameter $\ell$ being AdS radius. In AdS$_{d+1}$ spacetime,  the Ricci scalar curvature is related to the AdS radius $\ell$ through
\begin{equation}
\mathcal{R}=-\frac{d(d+1)}{\ell^2}\,.
\end{equation}
In global coordinates $\left(t, r, \theta_1, \theta_2, \ldots, \theta_{d-1}\right)$ covering the whole hyperboloid, we have the  following transformation
\begin{equation}\label{fjewi3pj4}
\begin{aligned}
z^0 & =\sqrt{\ell^2+r^2} \sin (t / \ell)\,, \\
z^1 & =r \cos \theta_1, \quad z^2=r \sin \theta_1 \cos \theta_2, \ldots, \\
z^{d-1} & =r \sin \theta_1 \cos \theta_2 \cdots \sin \theta_{d-2} \cos \theta_{d-1}\,, \\
z^d & =r \sin \theta_1 \sin \theta_2 \cdots \sin \theta_{d-2} \sin \theta_{d-1}\,, \\
z^{d+1} & =\sqrt{\ell^2+r^2} \cos (t / \ell)\, .
\end{aligned}
\end{equation}
Therefore, we obtain the familiar AdS${}_{d+1}$ metric
\begin{equation}\label{jgfi34pi}
\mathrm{d} s^2=-\left(1+r^2 / \ell^2\right) \mathrm{d} t^2+\frac{\mathrm{d} r^2}{1+r^2 / \ell^2}+r^2 \mathrm{d} \Omega_{d-1}^2
\end{equation}
with $\mathrm{d} \Omega_{d-1}^2$ denoting the line element for unit $(d-1)$-dimensional sphere. The pullback of the Wightman function
to a detector’s worldline is defined as the the vacuum expectation value\begin{equation}
  {\mathcal{W}}(\tau,\tau'):= \bra{0}\hat{\phi}(\rm{x}(\tau))\hat{\phi}(\rm{x}(\tau'))\ket{0}\,,
    \label{e.O1}
\end{equation}
satisfying the Klein-Gordon wave equation 
\begin{equation}
\left( \square_{\rm{x}} + m^2 + \xi \mathcal{R} \right){\mathcal{W}}(\rm{x},\rm{x}') = 0
\label{eq:Wightman_eq}\,,
\end{equation}
where $\square_{\rm{x}} $ denotes the  d’Alembertian with respect to $\rm{x}$, $m$ is the mass of the scalar field (which in our case is 0), $\mathcal{R}$ is the Ricci scalar, and $\xi$ is a conformal coupling constant.
Consider a $(d+1)$-dimensional AdS spacetime conformally coupled with a massless real scalar field with transparent boundary conditions, the Wightman function takes the  form \cite{Decanini:2005gt,Decanini:2005eg,Jennings:2010vk,saharian2020quantum,Smith:2017vle}
\begin{equation}\label{www}
   {\mathcal{W}}(\tau, \tau') = \frac{\Gamma(d-1)}{(4\pi)^{d/2} \Gamma(\frac{d}{2})}\frac{1}{ {\boldsymbol{\sigma}(\rm{x},\rm{x}')}^{\frac{d-1}{2}}}\,.
\end{equation}
Here, $\Gamma$ represents the Gamma function, $\boldsymbol{\sigma}(\rm{x},\rm{x}')$ denotes the  squared
geodesic distance between two spacetime points. Consider a static detector in AdS$_{d+1}$, whose trajectory is described by
\begin{equation}
\mathrm{x}(\tau) = \{ t = \frac{\tau \ell}{\gamma}, \; r = R, \; \theta_i = \Theta_i\}\,, 
\end{equation}
where $\gamma = \sqrt{\ell^2+R^2}$, $R$ and $\Theta_i$ denote fixed spatial coordinates. In global coordinates, the squared geodesic distance between $\rm{x}$  and $\rm{x}' $ in the embedding space is 
\begin{equation}\label{jfi38kk9i}
\begin{aligned}
\boldsymbol{\sigma}(\rm{x},\rm{x}') =&-  (z^0\left(t, r, \theta_1, \theta_2, \ldots, \theta_{d-1}\right)-z^0\left(t', r, \theta_1, \theta_2, \ldots, \theta_{d-1}\right))^2 \\
  &+ (z^1\left(t, r, \theta_1, \theta_2, \ldots, \theta_{d-1}\right)-z^1\left(t', r, \theta_1, \theta_2, \ldots, \theta_{d-1}\right))^2 +\cdots \\
  &+ (z^d\left(t, r, \theta_1, \theta_2, \ldots, \theta_{d-1}\right)-z^d\left(t', r, \theta_1, \theta_2, \ldots, \theta_{d-1}\right))^2 \\
    &- (z^{d+1}\left(t, r, \theta_1, \theta_2, \ldots, \theta_{d-1}\right)-z^{d+1}\left(t', r, \theta_1, \theta_2, \ldots, \theta_{d-1}\right))^2 \\
    =&2 \gamma^2\left(\cos \left(\frac{\tau-\tau'}{\gamma}\right)-1\right)\,.
    \end{aligned}
\end{equation}
Substituting the squared geodesic
distance \eqref{jfi38kk9i} into \eqref{www}, we obtain the Wightman function pulled back onto the world-line of a static detector in AdS${}_{d+1}$ $(d\geq 2$; see \cite{Pitelli:2021oil,Louko:2014aba} for the special case
$ d=1)$ spacetime
\begin{equation}\label{WF}
{\mathcal{W}}(\tau, \tau') = \frac{\Gamma(d-1)}{(4\pi)^{d/2} \Gamma(\frac{d}{2})} \left[2i\gamma \sin \left(\frac{\tau -\tau'-i\epsilon}{2 \gamma}\right)\right]^{1-d}\,,
\end{equation}
where $\epsilon$ is  a regularization parameter, ensuring that the poles of the integrand we will encounter later are displaced from the real axis by a small imaginary part. As a crosscheck, this Wightman function  reproduces the AdS${}_3$ result obtained in \cite{Henderson:2018lcy}. This can also recover the $d+1$-dimensional Minkowski spacetime Wightman function \cite{Nystrom:2024oeq} in the large $\gamma$ limit.

\subsection {Analytic derivation of excitation probability $q$}
The element $q$ is given by
\begin{align}
    q =&\frac{ \lambda^2 }{2}\int_{-\infty}^{\infty} \mathrm{d}\tau\, \int_{-\infty}^{\infty} \mathrm{d}\tau' \chi(\tau)\chi(\tau')e^{-i\Omega_1(\tau-\tau')}{\mathcal{W}}(\tau,\tau')\nonumber\\
    =& \frac{\lambda^2}{2} 
    \int_{-\infty}^{\infty}\mathrm{d}\tau \, \int_{-\infty}^{\infty} \mathrm{d}\tau' \, 
    e^{-(\tau^2 + \tau'^2)/\sigma^2} e^{-i \Omega_1 (\tau - \tau')}{\mathcal{W}}(\tau,\tau')\,. 
\end{align}
Note that on the worldline of a  static detector, the Wightman functions depend only on the proper time difference, i.e., ${\mathcal{W}}(\tau,\tau')={\mathcal{W}}(\tau-\tau')$. Now let $u=\tau+\tau'$ and $s=\tau-\tau'$. The modulus of the Jacobian determinant of this variable transformation is $1/2$, so $\mathrm{d}\tau \mathrm{d}\tau'=\mathrm{d}u \mathrm{d}s/2$. Then $q$ is equal to 
\begin{align}\label{q38}
    q =& \frac{\lambda^2}{4}\int_{-\infty}^{\infty}  \mathrm{d}u\; \int_{-\infty}^{\infty} \mathrm{d}s\; e^{-(u^2+s^2)/2\sigma^2}e^{-i\Omega s}{\mathcal{W}}(s)\nonumber\\
    =& \frac{\lambda^2}{2}\sigma\sqrt{\frac{\pi}{2}} \int_{-\infty}^{\infty}\mathrm{d}s\; e^{-s^2/2\sigma^2}e^{-i\Omega s}{\mathcal{W}}(s)\,.
\end{align}
Here we let $\Omega_1=\Omega$ for simplicity. Substituting the regularized Wightman function \eqref{WF} into our expression for $q$ \eqref{q38}, 
we obtain
\begin{align}\label{qq1}
    q=& \frac{\lambda^2\sigma\sqrt{\pi/2} \gamma^{1-d}\Gamma(d-1)}{2(4\pi)^{d/2} \Gamma(\frac{d}{2})}\int_{-\infty}^{\infty} \mathrm{d}s\; e^{-s^2/2\sigma^2}e^{-i\Omega s}\left[2i\sin \left(\frac{s}{2\gamma}-i\epsilon\right)\right]^{1-d},
\end{align}
where we have absorbed the positive factor $1/2\gamma$ into $\epsilon$. Let us exploit the fact that $\sin(s)=(\exp(is)-\exp(-is))/2i$ and then use the generalized binomial theorem to find an explicit expression for $q$. First, we note that
\begin{align}\label{factout}
    2i\sin\left(\frac{s}{2\gamma} -i\epsilon \right)=&\; e^{i(\frac{s}{2\gamma}-i\epsilon)}-e^{-i(\frac{s}{2\gamma}-i\epsilon)}=e^{i(\frac{s}{2\gamma}-i\epsilon)}\left[1-e^{-2i(\frac{s}{2\gamma}-i\epsilon)}\right],
\end{align}
so we have
\begin{align}
\left[2i\sin\left(\frac{s}{2\gamma} -i\epsilon \right)\right]^{1-d}=e^{(1-d)i(\frac{s}{2\gamma}-i\epsilon)}\left[1-e^{-2i(\frac{s}{2\gamma}-i\epsilon)}\right]^{1-d}.
\end{align}
 Note that we are working with $d\geq2$, so the exponent $1-d$ is negative. We now turn to a result from the generalized binomial theorem
\begin{align}
    \frac{1}{(1-x)^{p}}=&\sum_{n=0}^{\infty}\binom{p+n-1}{n}x^{n}.
\end{align}
A key point is that in \eqref{factout}  the choice as to which exponential to factor out is not arbitrary (this treatment would not work if we had  factored out $e^{-i(\frac{s}{2\gamma}-i\epsilon)}$ instead). Because the binomial theorem expansion of $(1-x)^{-p}$ only converges for $|x|<1$, which is ensured here by our positive regularizing parameter $\epsilon$. Here, we have $p=d-1$, therefore
\begin{align}\label{trick}
    \left[2i\sin\left(\frac{s}{2\gamma} -i\epsilon \right)\right]^{1-d}=&\;e^{(1-d)i(\frac{s}{2\gamma}-i\epsilon)}\sum_{n=0}^{\infty}\binom{d+n-2}{n}e^{-2in(\frac{s}{2\gamma}-i\epsilon)}\nonumber\\
    =&\sum_{n=0}^{\infty}\binom{d+n-2}{n}e^{-i(2n+d-1)(\frac{s}{2\gamma}-i\epsilon)}\,.
\end{align}
Substituting this sum into our expression for $q$ in \eqref{qq1} yields
\begin{align}\label{q44}
q=&\frac{\lambda^2\sigma\sqrt{\pi/2}\gamma^{1-d}\Gamma(d-1)}{2(4\pi)^{d/2} \Gamma(\frac{d}{2})}\sum_{n=0}^{\infty}\binom{d+n-2}{n}e^{-(2n+d-1)\epsilon}\int_{-\infty}^{\infty} \mathrm{d}s\; e^{-s^2/2\sigma^2}e^{-i\Omega s}e^{-i(2n+d-1)s/2\gamma}.
\end{align}
Now we are left with a sum over Fourier transforms of a Gaussian, which can be evaluated explicitly
\begin{align}
   \int_{-\infty}^{\infty}\mathrm{d}s \; e^{-s^2/2\sigma^2}e^{-i \tilde{\Omega} s}=&\sqrt{2\pi}\sigma e^{-\tilde{\Omega}^2\sigma^2/2}.
\end{align}
We define $\Omega_n\equiv\frac{2n+d-1}{2\gamma}$, thus in our case $\tilde{\Omega}=\Omega+\Omega_n$, which can be understood as an effective  energy/frequency modified by the mode number $n$, and spacetime structure parameters $\gamma$ and $d$. Substituting the above integration result into our expression for $q$ \eqref{q44} gives 
\begin{align}
    q=&\frac{\lambda^2\sigma^{2}\pi\gamma^{1-d}\Gamma(d-1)}{2(4\pi)^{d/2} \Gamma(\frac{d}{2})}\sum_{n=0}^{\infty}e^{-(2n+d-1)\epsilon}\binom{d+n-2}{n}e^{-\frac{\sigma^{2}}{2}\left(\Omega+\frac{2n+d-1}{2\gamma}\right)^{2}}.
\end{align}
Taking the limit of $\epsilon\rightarrow0^{+}$, we get
\begin{align}
    q=&\frac{\lambda^2\sigma^{2}\pi\gamma^{1-d}\Gamma(d-1)}{2(4\pi)^{d/2} \Gamma(\frac{d}{2})}\sum_{n=0}^{\infty}\binom{d+n-2}{n}e^{-\frac{\sigma^{2}}{2}\left(\Omega+\frac{2n+d-1}{2\gamma}\right)^{2}}.
\end{align}
Finally, we can write the binomial coefficient in terms of the $\Gamma$ function to make things neater
\begin{align}
q=2\alpha\sum_{n=0}^{\infty}\frac{\Gamma(d+n-1)}{\Gamma(n+1)}e^{-\frac{\sigma^{2}}{2}\left(\Omega+\Omega_n\right)^{2}}\,,
\end{align}
where $\alpha\equiv\frac{\lambda^2\sigma^{2}\pi\gamma^{1-d}}{4(4\pi)^{d/2} \Gamma(\frac{d}{2})}$, $\Omega_n\equiv\frac{2n+d-1}{2\gamma}$, $\gamma\equiv\sqrt{\ell^2+R^2}$,  and $\Gamma$ represents the Gamma function.
 This expression agrees with numerical integration results. Furthermore, for $\text{AdS}_{4}$, in the limit of $\ell\rightarrow\infty$, the result agrees with the excitation probability of an inertial UDW detector in $(3+1)$-dimensional Minkowski spacetime \cite{Nystrom:2024oeq}, as it should.

\subsection{Analytic derivation of off-diagonal term $\beta$}
The element $\beta$  is given by
\begin{align}
    \beta 
    =& -\frac{\lambda^2}{4}  \int_{-\infty}^{\infty} \mathrm{d} \tau \mathrm{d} \tau' \bigg[ \Theta(\tau-\tau') e^{-(\tau^2 + \tau'^2)/\sigma^2} e^{i\Omega_2 \tau + i \Omega_1 \tau'}{\mathcal{W}}(\tau,\tau') + \Theta(\tau'-\tau) e^{-(\tau^2 + \tau'^2)/\sigma^2} e^{i\Omega_1 \tau + i \Omega_2 \tau'}{\mathcal{W}}(\tau',\tau) \bigg].
\end{align}
For a three-level system with equal energy gaps between adjacent levels $\Omega_{1}=\Omega_{2}=\Omega$, the expression for $\beta$ can be further simplified to
\begin{align}
    \beta 
    =& -\frac{\lambda^2}{4}  \int_{-\infty}^{\infty} \mathrm{d} \tau \int_{-\infty}^{\infty} \mathrm{d} \tau' \bigg[ e^{-(\tau^2 + \tau'^2)/\sigma^2} e^{i\Omega(\tau +\tau')} \left(\Theta(\tau-\tau'){\mathcal{W}}(\tau,\tau') + \Theta(\tau'-\tau){\mathcal{W}}(\tau',\tau)\right) \bigg].
\end{align}
Once again, we exploit the fact that the Wightman functions are functions of $\tau-\tau'$ for a static trajectory, and make a variable switch to $u=\tau+\tau'$ and $s=\tau-\tau'$.  The integral becomes
\begin{align}
    \beta=&-\frac{\lambda^2}{8} \int_{-\infty}^{\infty} \mathrm{d}u\; \int_{-\infty}^{\infty} \mathrm{d}s\; \bigg[ e^{-(u^2 + s^2)/2\sigma^2} e^{i\Omega u} \left(\Theta(s){\mathcal{W}}(s) + \Theta(-s){\mathcal{W}}(-s)\right) \bigg]\,.
\end{align}
Carrying out the integral over $u$, we obtain
\begin{align}\label{betaexp}
    \beta=&-\frac{\lambda^2\sqrt{2\pi}\sigma}{8}e^{-\sigma ^2 \Omega ^2/2} \int_{-\infty}^{\infty} \mathrm{d}s\; e^{-s^2/2\sigma^2} \left(\Theta(s){\mathcal{W}}(s) + \Theta(-s){\mathcal{W}}(-s)\right)\nonumber\\
    =&-\frac{\lambda^2\sqrt{2\pi}\sigma}{4}e^{-\sigma ^2 \Omega ^2/2} \int_{0}^{\infty} \mathrm{d}s\; e^{-s^2/2\sigma^2}{\mathcal{W}}(s)\,.
\end{align}
Substituting the Wightman function into the expression for $\beta$, we get
\begin{align}\label{int}
    \beta =& -\frac{\lambda^{2}\sigma\sqrt{2 \pi }\gamma^{1-d}\Gamma(d-1)}{4(4\pi)^{d/2} \Gamma(\frac{d}{2})} e^{- \sigma ^2 \Omega ^2/2}\int_{0}^{\infty} \mathrm{d}s\; e^{-s^2/2\sigma^2}\left[2i \sin \left(\frac{s}{2\gamma}-i\epsilon\right)\right]^{1-d}.
\end{align}
Now we can repeat the binomial trick \eqref{trick}, thus the integral in \eqref{int} becomes
\begin{align}
    &\int_{0}^{\infty} \mathrm{d}s\; e^{-s^2/2\sigma^2}\left[2i \sin \left(\frac{s}{2\gamma}-i\epsilon\right)\right]^{1-d}\nonumber\\
    =&\sum_{n=0}^{\infty}\binom{d+n-2}{n}e^{-(2n+d-1)\epsilon}\int_{0}^{\infty} \mathrm{d}s\; e^{-s^2/2\sigma^2}e^{-i(2n+d-1)s/2\gamma}\nonumber\\
    =&\sqrt{\frac{\pi}{2}}\sigma\sum_{n=0}^{\infty}\binom{d+n-2}{n}e^{-(2n+d-1)\epsilon} e^{-\frac{\sigma^{2}}{2}\left(\frac{2n+d-1}{2\gamma}\right)^{2}}\left[1-i\;\text{erfi} \left[\frac{\sigma}{\sqrt{2}}\frac{(2n+d-1)}{2\gamma}\right]\right].
\end{align}
By substituting the above integral result into \eqref{int} and taking the regulator $\epsilon\to0^{+}$, we arrive at the final expression for the off-diagonal element $\beta$,
\begin{align}\label{beta51}
\beta =-\alpha \sum_{n=0}^{\infty}\frac{\Gamma(d+n-1)}{\Gamma(n+1)}e^{-\frac{\sigma^{2}}{2}\left(\Omega^2+\Omega_n^2\right)}\left[1-i\;\text{erfi} \left[\frac{\sigma}{\sqrt{2}}\Omega_n\right]\right].
\end{align}
where $\alpha\equiv\frac{\lambda^2\sigma^{2}\pi\gamma^{1-d}}{4(4\pi)^{d/2} \Gamma(\frac{d}{2})}$, $\Omega_n\equiv\frac{2n+d-1}{2\gamma}$, $\gamma\equiv\sqrt{\ell^2+R^2}$, and erfi$(z)$ is the imaginary error function defined by $\text{erf}(iz)/i$. Since this is effectively the correlation term of two UDW detectors in the coincident limit, we have to deal with the problem of the UV divergence. Both numerically, and symbollically, our expression for $\beta$ diverges. The trick used by the authors of \cite{Nystrom:2024oeq}, which is to analytically continue the explicit result of the integral from its domain of strict validity to $d=3$, will not work, since we are left with an infinite sum  with no closed form expression. Nevertheless, the structure of the expression in \eqref{beta51} allows for a clear separation into two terms, one term is responsible for the divergence and the other term remains finite. Note that for large $n$, the Gamma functions $\frac{\Gamma(d+n-1)}{\Gamma(n+1)}\sim n^{d-2}$, so due to the exponential decay of the Gaussian, the product of the Gamma functions and the Gaussian converges. While the erfi function  behaves like $\text{erfi}(n)\sim-i+\frac{e^{n^2}}{\sqrt{\pi} n} $, the exponential  suppression  of the Gaussian is cancelled with the exponentially divergent term due to the erfi function, thus the term with the erfi function diverges. So the divergent term is given by
\begin{align}
    \beta_{\text{div}}=-\alpha \sum_{n=0}^{\infty}\frac{\Gamma(d+n-1)}{\Gamma(n+1)}e^{-\frac{\sigma^{2}}{2}\left(\Omega^2+\Omega_n^2\right)}\left[-i\;\text{erfi} \left[\frac{\sigma}{\sqrt{2}}\Omega_n\right]\right].
\end{align}

Therefore, we can define a renormalized  $\beta$
\begin{align}
    \beta_{\text{ren}} =    \beta -    \beta_{\text{div}} =-\alpha \sum_{n=0}^{\infty}\frac{\Gamma(d+n-1)}{\Gamma(n+1)}e^{-\frac{\sigma^{2}}{2}\left(\Omega^2+\Omega_n^2\right)}\,,
\end{align}
 where $\alpha\equiv\frac{\lambda^2\sigma^{2}\pi\gamma^{1-d}}{4(4\pi)^{d/2} \Gamma(\frac{d}{2})}$, $\Omega_n\equiv\frac{2n+d-1}{2\gamma}$, and $\gamma\equiv\sqrt{\ell^2+R^2}$. Moreover, the troublesome $\text{erfi}$ function, which leads to the divergence, comes  from the fact that our integral is from $s=0$ to $s=\infty$, instead of over the entire real line. When it came to $q$, the contributions from various poles (which are symmetrically distributed about $s=0$) cancelled each other out, and we were left with a finite result. Therefore, an alternative way is to extend the integral to the entire real line and divide by 2, which effectively eliminates the 
erfi function. Also note that the expression can be further simplified using the Pochhammer symbol, $(x)_{y}=\frac{\Gamma(x+y)}{\Gamma(x)}$, thus the Gamma functions become $  \frac{\Gamma(d+n-1)}{\Gamma(n+1)}=(n+1)_{d-2}= \prod_{k=1}^{d-2} (n + k)$. However, we shall leave the formulas in terms of Gamma functions to enhance readability. Again, this expression agrees
with numerical testing and recovers the Minkowskian results \cite{Nystrom:2024oeq}.  In what follows and in main text, we shall denote the renormalized $\beta_{\text{ren}}$  by $\beta$ for simplicity.

\subsection{Closed-form formulas for $q$  in large $\gamma$ regime}

Although we have obtained the analytic expressions for $q$ and $\beta$ in terms of infinite series, it is of interest to explore  whether they admit closed-form formula. Start with $q$ 
\begin{align}
    q(\gamma)=\frac{\lambda^2\sigma^{2}\pi\gamma^{1-d}}{2(4\pi)^{d/2} \Gamma(\frac{d}{2})}\sum_{n=0}^{\infty}\frac{\Gamma(d+n-1)}{\Gamma(n+1)}e^{-\frac{\sigma^{2}}{2}\left(\Omega+\frac{2n+d-1}{2\gamma}\right)^{2}}.
\end{align}
Let $\Omega_d\equiv  \frac{d-1}{2\gamma}$, $\alpha\equiv\frac{\lambda^2\sigma^{2}\pi\gamma^{1-d}}{4(4\pi)^{d/2} \Gamma(\frac{d}{2})}$,  define $x_{n}=\frac{n}{\gamma}$, so then $\Delta x_{n}=\frac{1}{\gamma}$. Multiplying and dividing the sum by $\gamma$, we obtain
\begin{align}
    q(\gamma)=2\gamma \alpha\sum_{n=0}^{\infty}\frac{\Gamma(d+\gamma x_n-1)}{\Gamma(\gamma x_n+1)}e^{-\frac{\sigma^{2}}{2}\left(\Omega+\Omega_d+x_{n}\right)^{2}}\Delta x_n\,.
\end{align}
We recognize this as a Riemann sum, that in the limit of large $\gamma$, approaches an integral,
\begin{align}
  q(\gamma)=  2\gamma \alpha\sum_{n=0}^{\infty}\frac{\Gamma(d+\gamma x_n-1)}{\Gamma(\gamma x_n+1)}e^{-\frac{\sigma^{2}}{2}\left(\Omega+\Omega_d+x_{n}\right)^{2}}\Delta x_n\longrightarrow2\gamma \alpha\int_{0}^{\infty}\mathrm{d}x\;\frac{\Gamma(\gamma x+d-1)}{\Gamma(\gamma x+1)}e^{-\frac{\sigma^{2}}{2}\left(\Omega+\Omega_d+x\right)^{2}}\,.
\end{align}
Applying the trick  $  \frac{\Gamma(\gamma x+d-1)}{\Gamma(\gamma x+1)}= \prod_{k=1}^{d-2} (\gamma x + k)=\sum_{n=0}^{d-2}\mathcal{C}_n^{(d)}  x^n$, where $ \mathcal{C}_n^{(d)}$
 represents the coefficient of  $x^n$ in $ \prod_{k=1}^{d-2} (\gamma x + k)
$,  this integral can be rewritten as
\begin{align}
2\gamma \alpha\int_{0}^{\infty}\mathrm{d}x\;\frac{\Gamma(\gamma x+d-1)}{\Gamma(\gamma x+1)}e^{-\frac{\sigma^{2}}{2}\left(\Omega+\Omega_d+x\right)^{2}}=2\gamma \alpha\sum_{n=0}^{d-2}\mathcal{C}_n^{(d)} \int_{0}^{\infty}\mathrm{d}x\;x^n e^{-\frac{\sigma^{2}}{2}\left(\Omega+\Omega_d+x\right)^{2}}\,,
\end{align}
which yields a closed-form representation for $q$, denoted by $q_c$,
\begin{align}\label{closeq}
   q_c= 2\gamma \alpha\sum_{n=0}^{d-2} 
\zeta_n\left[\Gamma \left(\frac{n+1}{2}\right)  F\left(-\frac{n}{2};\frac{1}{2};\frac{- \sigma ^2 (\Omega+\Omega_d )^2}{2}\right)-\sqrt{2} \sigma(\Omega+\Omega_d ) \Gamma \left(\frac{n+2}{2}\right)  F\left(\frac{1-n}{2};\frac{3}{2};\frac{-\sigma ^2 (\Omega+\Omega_d )^2}{2}\right)\right]\,,
\end{align}
where $F(a;b;c)$ is the Kummer confluent hypergeometric function, $d$ is the spatial dimension, and $\zeta_n= 2^{\frac{n-1}{2}}\sigma ^{-n-1}\mathcal{C}_n^{(d)} $. In Table \ref{tab:q-dimensions}, we present the simplified expressions fo $q_c$  obtained from \eqref{closeq} in large $\gamma$ regime for  dimensions $d = 2, 3, 4$.

\begin{table}[htbp]
\caption{Closed-form expressions for the detector excitation probability $q_c$ in the large-$\gamma$ regime for dimensions $d = 2, 3, 4$.}
\label{tab:q-dimensions}
\centering
\renewcommand{\arraystretch}{1.4}
\begin{ruledtabular}
\begin{tabular}{cl}
\text{Dimension} & \text{Expression for $q_c$} \\
\hline
$d = 2$ &
$\displaystyle \frac{1}{8} \sqrt{\frac{\pi}{2}} \lambda^2 \sigma\ \mathrm{erfc}\left( \frac{\sigma(1 + 2\gamma \Omega)}{2\sqrt{2}\gamma} \right)$ \\
$d = 3$ &
$\displaystyle \frac{\lambda^2}{16\pi} \left( 2 e^{- \frac{(\sigma + \gamma \sigma \Omega)^2}{2 \gamma^2}} - \sqrt{2\pi} \sigma \Omega\ \mathrm{erfc}\left( \frac{\sigma(1 + \gamma \Omega)}{\sqrt{2}\gamma} \right) \right)$ \\
$d = 4$ &
$\displaystyle \frac{\lambda^2}{256\pi \gamma^2 \sigma} \left( 4\gamma \sigma(3 - 2\gamma \Omega) e^{ -\frac{\sigma^2 (3 + 2\gamma \Omega)^2}{8\gamma^2} } + \sqrt{2\pi} \left( 4\gamma^2(1 + \sigma^2 \Omega^2) - \sigma^2 \right)\ \mathrm{erfc}\left( \frac{\sigma(3 + 2\gamma \Omega)}{2\sqrt{2}\gamma} \right) \right)$ \\
\end{tabular}
\end{ruledtabular}
\end{table}

While we can numerically validate that the limit of $\gamma\rightarrow\infty$ recovers the results of Minkowski spacetime, it is helpful to see this analytically. In the limit of large $\gamma$, the Gamma functions  behave like $ \frac{\Gamma(\gamma x+d-1)}{\Gamma(\gamma x+1)}\sim \gamma^{d-2}x^{d-2}$, and the factor of $\gamma^{d-2}$ cancels out the constant factor $\gamma^{2-d}$ in front of our sum. The term $\Omega_d=\frac{d-1}{2\gamma}$ in the exponential becomes zero in the limit of large $\gamma$. Replacing the Riemann sum with the exact integral then shows us
\begin{align}
    \lim_{\gamma\rightarrow\infty} q(\gamma)=\frac{\lambda^2\sigma^{2}\pi}{2(4\pi)^{d/2} \Gamma(\frac{d}{2})}\int_{0}^{\infty}\mathrm{d}x\; x^{d-2}e^{-\frac{\sigma^{2}}{2}\left(\Omega+x\right)^{2}}.
\end{align}
This integral can be evaluated exactly, so we get the excitation probability of a detector in $(d+1)$-dimensional Minkowski spacetime,
\begin{align}
  q_{M_{d+1}}=  \lim_{\gamma\rightarrow\infty}q(\gamma)
    =&\frac{\lambda^2\Omega\sigma^{4-d}\Gamma(d-1)\pi}{2(8\pi)^{d/2} \Gamma(\frac{d}{2})} e^{-\frac{1}{2} \sigma ^2 \Omega ^2} U\left(\frac{d}{2},\frac{3}{2},\frac{\sigma ^2 \Omega ^2}{2}\right),
\end{align}
where $U(a,b,c)$ is the Tricomi confluent hypergeometric function. Consider a special case when $d=3$,  we get
\begin{align}
    q_{M_{3+1}}=\frac{\lambda^{2}\sigma  \Omega}{16 \sqrt{2} \pi }\Gamma \left(-\frac{1}{2},\frac{\sigma ^2 \Omega ^2}{2}\right)=\frac{\lambda ^2 \left(2 e^{-\frac{1}{2} \sigma ^2 \Omega ^2}-\sqrt{2 \pi } \sigma  \Omega  \text{erfc}\left(\frac{\sigma  \Omega }{\sqrt{2}}\right)\right)}{16 \pi }\,.
\end{align}
Here, $\Gamma(s,z)$ is the upper incomplete $\Gamma$ function, erfc$(z)$ is the complementary error function. Thus we have tested this against the known expression for $q$ in four-dimensional  Minkowski spacetime  \cite{Nystrom:2024oeq}.  As we can also see from Table \ref{tab:q-dimensions}, our  $q_c$ recovers the Minkowski result, as $\gamma\to \infty$.

\subsection{Closed-form formulas for $\beta$ in large $\gamma$ regime}

We now turn to find the closed-form for $\beta$. 
Once more, we define $x_{n}=\frac{n}{\gamma}$,  $\Omega_d=  \frac{d-1}{2\gamma}$, and repeat the previous procedure to replace the sum, in the limit of large $\gamma$, with an integral
\begin{align}
  \beta=-\alpha  &\sum_{n=0}^{\infty}\frac{\Gamma(d+\gamma x_{n}-1)}{\Gamma(\gamma x_{n}+1)}e^{-\frac{\sigma^{2}}{2}\left(\Omega^2+(\Omega_d+x)^2\right)}\Delta x_{n}\longrightarrow-\alpha\int_{0}^{\infty}\mathrm{d}x\;\frac{\Gamma(\gamma x+d-1)}{\Gamma(\gamma x+1)}e^{-\frac{\sigma^{2}}{2}\left(\Omega^2+(\Omega_d+x)^2\right)}.
\end{align}
Again, we apply the trick  $  \frac{\Gamma(\gamma x+d-1)}{\Gamma(\gamma x+1)}= \prod_{k=1}^{d-2} (\gamma x + k)=\sum_{n=0}^{d-2}\mathcal{C}_n^{(d)}  x^n$and thereby obtain a closed-form expression for $\beta$, denoted by $\beta_c$,
\begin{align}\label{closeb}
   \beta_c= -\gamma \alpha\sum_{n=0}^{d-2} 
\zeta_n e^{-\frac{1}{2} \sigma ^2 \left(\Omega^2+\Omega_d ^2\right)} \left[\Gamma \left(\frac{n+1}{2}\right)  F\left(\frac{n+1}{2};\frac{1}{2};\frac{\Omega_d^2 \sigma ^2}{2}\right)-\sqrt{2} \sigma  \Omega_d \Gamma \left(\frac{n+2}{2}\right)F\left(\frac{n}{2}+1;\frac{3}{2};\frac{\Omega_d^2 \sigma ^2}{2}\right)\right]\,,
\end{align}
where $F(a;b;c)$ is the Kummer confluent hypergeometric function, and $\zeta_n= 2^{\frac{n-1}{2}}\sigma ^{-n-1}\mathcal{C}_n^{(d)} $. Table~\ref{tab:b-dimensions} lists the simplified expressions for $\beta_c$  derived from Eq.~\eqref{closeb} for $d = 2,\ 3,\ 4$, in the regime where $\gamma$ is large.

\begin{table}[htbp]
\caption{Closed-form expressions for the detector coherence $\beta_c$ in the large-$\gamma$ regime for dimensions $d = 2, 3, 4$.}
\label{tab:b-dimensions}
\centering
\renewcommand{\arraystretch}{1.4}
\begin{ruledtabular}
\begin{tabular}{cl}
\text{Dimension} & \text{Expression for $\beta_c$} \\
\hline
$d = 2$ &
$\displaystyle -\frac{1}{16} \sqrt{\frac{\pi }{2}} \lambda ^2 \sigma  e^{-\frac{1}{2} \sigma ^2 \Omega ^2} \text{erfc}\left(\frac{\sigma }{2 \sqrt{2} \gamma }\right)$  \\
$d = 3$ &
$\displaystyle -\frac{1}{16 \pi }\lambda ^2 e^{-\frac{1}{2} \sigma ^2 \left(\frac{1}{\gamma ^2}+\Omega ^2\right)}$ \\
$d = 4$ &
$\displaystyle \frac{1}{512 \pi  \gamma ^2 \sigma }\lambda ^2 e^{-\frac{1}{2} \sigma ^2 \left(\frac{9}{4 \gamma ^2}+\Omega ^2\right)} \left(\sqrt{2 \pi } e^{\frac{9 \sigma ^2}{8 \gamma ^2}} \left(\sigma ^2-4 \gamma ^2\right) \text{erfc}\left(\frac{3 \sigma }{2 \sqrt{2} \gamma }\right)-12 \gamma  \sigma \right)$ \\
\end{tabular}
\end{ruledtabular}
\end{table}

Taking the limit of $\gamma\rightarrow\infty$, and substituting back into the expression for $\beta$, we get the (renormalized) expression for $\beta$ in $d+1$-dimensional Minkowski spacetime
\begin{align}
    \beta_{M_{d+1}}=\lim_{\gamma\rightarrow\infty}\beta(\gamma)=&-\frac{\lambda^{2}\sigma^{2} \pi}{4(4\pi)^{d/2} \Gamma(\frac{d}{2})} \int_{0}^{\infty}\mathrm{d}x\; x^{d-2}e^{-\sigma^{2}(\Omega^2+x^{2})/2}\nonumber\\
    =&-\frac{\lambda^{2}2^{-\frac{d+7}{2}} \pi ^{1-\frac{d}{2}} \sigma ^{3-d} \Gamma \left(\frac{d-1}{2}\right)}{\Gamma \left(\frac{d}{2}\right)}e^{- \sigma ^2 \Omega ^2/2}.
\end{align}
We can substitute in $d=3$ and get
\begin{align}
    \beta_{M_{3+1}}=-\frac{\lambda^{2}}{16\pi}e^{-\sigma^{2}\Omega^{2}/2},
\end{align}
which agrees with the  four-dimensional Minkowski result  \cite{Nystrom:2024oeq}.

At this stage, one may wonder whether the integral remains a valid approximation of the series across an arbitrary $\gamma$ regime, and what the difference between the two actually is. Fortunately, the Euler--Maclaurin formula provides a precise bridge between sums and integrals,
\begin{equation}\label{EM}
\sum_{n=a}^{b} f(n) =\int_a^b f(x) \, \mathrm{d}x+R_m\,.
\end{equation}
Here, the remainder term between the sum and the integral is
\begin{align}
    R_m =   \frac{f(a) + f(b)}{2}
+ \sum_{k=1}^{m} \frac{B_{2k}}{(2k)!} \left( f^{(2k-1)}(b) - f^{(2k-1)}(a) \right)+ \frac{(-1)^{m+1}}{(2m)!} \int_a^b B_{2m}(x - \lfloor x \rfloor)\, f^{(2m)}(x)\, \mathrm{d}x \,,
\end{align}
where \( B_{2k} \) are Bernoulli numbers, \( f^{(2k-1)}(x) \) is the \(2k-1\)-th derivative of \( f(x) \),  \( B_{2m}(x - \lfloor x \rfloor) \) is the \(2m\)-th Bernoulli polynomial, and \( \lfloor x \rfloor \) denotes the floor function. Notably, Eq.~\eqref{EM} remains valid for arbitrary $\gamma$.  In our case, by replacing the infinite series expressions for $q$ and $\beta$ with their integral counterparts, one can easily verify $\int_a^b f(x) \, \mathrm{d}x$ are exactly the closed-form expressions $q_c$ and $\beta_c$ obtained previously. Moreover, the remainder term $R_m$ vanishes in the limit $\gamma \to \infty$, justifying the validity of the integral approximation in this regime.  For small values of $\gamma \sim \mathcal{O}(1)$, it is sufficient to set $m = 2$ in the remainder estimate: the term $\frac{(-1)^{m+1}}{(2m)!} \int_a^b B_{2m}(x - \lfloor x \rfloor)\, f^{(2m)}(x)\, \mathrm{d}x$ is of the order of $10^{-6}$, indicating excellent agreement between the integral approximation (plus three other terms) and the original series. Hence,  even in small $\gamma$ regime, closed-form expressions for $q$ and $\beta$  can be achieved within the allowed error tolerance by including several remainder terms.

\end{document}